\documentclass[showpacs,preprint]{revtex4}
\usepackage{graphicx}
\usepackage{subfigure}
\usepackage{amssymb}
\usepackage{amsmath}
\usepackage{bm}
\usepackage{latexsym}
\usepackage{hyperref}

\begin {document}

\title{Non-local dynamics of Bell states in separate cavities}

\author{Jun Jing$^{1}$\footnote{Email address: jungen@shu.edu.cn},
Zhi-guo L\"{u}$^2$, Guo-hong Yang$^1$}

\affiliation{$^1$Department of Physics, Shanghai University,
Shanghai 200444, China\\ $^2$Department of Physics, Shanghai
Jiaotong University, Shanghai 200240, China}

\date{\today}

\begin{abstract}
We present non-local dynamics of Bell states in separate cavities.
It is demonstrated that (i) the entanglement damping speed will
saturate when the cavity leakage rate $\gamma\geq 0.4$; (ii) the
synchronism relationship between the fidelity and the concurrence
depends on the initial state; (iii) if the initial state is
$1/\sqrt{2}(|01\rangle+|10\rangle)$, the dynamics of entropy is
opposite to that of fidelity.

\end{abstract}

\maketitle

\section{Introduction}

In contrast with the extensively investigated static entanglement
\cite{Nielsen, Bennett, Horodecki, Werner, Greenberger, Dur},
dynamic entanglement under the influence of variant environments is
one of the most important and largely unexplored problems in the
field of quantum teleportation, quantum computation and quantum
communication \cite{Yu1, Diosi, Yu2, Carvalho}. It is not only
involved with the foundation of quantum mechanics, but also a
fundamental issue in creating, quantifying, controlling,
distributing and manipulating the entangled quantum bits, which are
composed of spin-$1/2$ atoms in different problems \cite{Nielsen,
Loss, Kane, Tanas}. An entangled system is in such a state that
cannot be factorized \cite{Shimony} in its Hilbert space. And the
most familiar and widely used examples are Bell states. The two
particles or atoms of spin-$1/2$ are correlated no matter how long
distance is between them. Generally, due to different kinds of
quantum reservoir, the entanglement degree between them vanishes
asymptotically. However, if the reservoir consists of, e.g., only
one or two electromagnetic field modes, then the entanglement may
decrease abruptly and non-smoothly to zero in a finite time
\cite{Yu2, Yu3, Ficek}, which is a new nonlocal decoherence called
entanglement sudden death (ESD). Therefore, demonstration of the
dynamics of Bell states \cite{Liang} would have profound
implications for understanding of the physics in the realization of
qubits in experiments. \\

So far in quantum optics experiments, Bell states can be generated
with trapped ions \cite{Sackett} and in cavity quantum
electrodynamics (CQED) \cite{Guo, Rau}, which has attracted much
attention. Based on cavity QED systems, schemes (to see Refs.
\cite{Cirac, Pell, Duan, Mancini, Serafini} and references therein)
have been proposed to implement quantum communications or engineer
entanglement between atoms in distant optical cavities. In most of
them, two separated cavities are connected via some channels, for
instance, an optical fiber \cite{Serafini}. And in a recent paper,
Yin and Li \cite{Yin} investigated a system consisting of two
single-mode cavities connected by an optical fiber and multiple
two-level atoms trapped in the cavities. They show that ideal
entangling can be deterministically realized between the distant
cavities. Besides, utilizing a system of two-atoms and two photon
modes, Masood and Miller \cite{Samina} used the Jaynes and Cummings
model \cite{JCM}, which is considered to be one of the most
appropriate models for exploiting the dynamics of entanglement
\cite{Bose, Zhou, Rendell, Isabel}, in the rotating wave
approximation to study entanglement of more than one atom with
vacuum. The photon modes in their model are uncoupled, however, the
leakage of cavities and the effect of temperature (Yet recently, in
other models the two-qubit entanglement dynamics for a
finite-temperature environment has been discussed in \cite{XFZhou,
FLLi}) are actually not considered thoroughly. \\

In this paper, we consider a quantum model with two identical
two-level atoms or pseudo-spins of $1/2$ (as an open subsystem with
qubits labelled $s_1$ and $s_2$) and two single-mode cavities
(labelled $1$ and $2$ correspondingly). The atom $s_j$ ($j=1$ or
$2$) is embedded in and coupled only with the cavity mode $j$, which
could be regarded as its bath or environment. The two cavities are
so far departed that there is no direct interaction between them as
well as the two atoms. Initially, the two qubits are prepared as a
most-entangled states (Bell states). The focus of interest is their
degrading quantum evolution, which are measured by the concurrence
\cite{Wootters1, Wootters2}, the fidelity \cite{Privman} and the
entropy exchange \cite{Benjamin, Benjamin2}. The calculations and
physical arguments will be carried out in two conditions: (i) there
is leakage of photons for the cavities, which are in the vacuum
states from the beginning; (ii) the cavities are so perfect that the
loss of photons from them could be neglected and the two single
modes are initialled in a thermal equilibrium state with the same
temperature. The rest of this paper is organized as following. In
Sec. \ref{Model} we begin with the model Hamiltonian and its
analysis derivation; and then we introduce the numerical calculation
procedure about the evolution of the reduced matrix for the
subsystem. Detailed results and discussions can be found in Sec.
\ref{discussion}. We will make a conclusion in Sec.
\ref{conclusion}.

\section{Model and Method}\label{Model}

The master equation for a two-level atom in a single-mode cavity
\cite{Scala}, as one of the two partitions in our model, can be
taken as
\begin{equation}\label{Lind}
i\frac{d\rho_j}{dt}=[H_j,\rho_j]+i\gamma_j\left(a_j\rho_ja^\dagger_j-
\frac{1}{2}a^\dagger_ja_j\rho_j-\frac{1}{2}\rho_ja^\dagger_ja_j\right).
\end{equation}
For density matrix $\rho_j$, $j$ refers to $s_1$ or $s_2$; for the
mode operator $a_j$ or $a^\dagger_j$, $j$ ($1$ or $2$) represents
the photon mode coupling with the corresponding atom. $\gamma_j$ is
the leakage rate of photons from the cavity $j$. $H_j$ describes the
Hamiltonian for a subsystem of one atom and one cavity ($j=1, 2$):
\begin{equation}\label{Hamj}
H_j=\frac{\omega_j}{2}\sigma^z_j+(1+\epsilon_j)\omega_ja^\dagger_ja_j+
g_j\omega_j(a^\dagger_j+a_j)\sigma^x_j.
\end{equation}
where $\omega_j$ is the energy level difference of atom $s_j$ in
cavity $j$. $\epsilon_j$ is the detuning parameter measuring the
deviation of the photon $j$ energy from $\omega_j$. $g_j$ is
introduced as another dimensionless parameter which suggests the
coupling strength between qubit $s_j$ and mode $j$. The $x$ and $z$
components of $\sigma$ are the well-known Pauli operator. The two
qubits are embedded in remote cavities without direct interaction.
Therefore the whole Hamiltonian for this two-atom-two-cavity problem
is
\begin{equation}
H=H_1+H_2.
\end{equation}

The whole state of the total system is assumed to be separable
before $t=0$, i.e.
\begin{eqnarray}
\rho(0)&=&\rho_S(0)\otimes\rho_b(0),\\
\rho_S(0)&=&|\psi(0)\rangle\langle\psi(0)|, \\
\rho_b(0)&=&\rho_{b1}(0)\otimes\rho_{b2}(0).
\end{eqnarray}
The initial state $|\psi(0)\rangle$ for the two qubits is one of the
Bell states. And the two cavities are in their (i) vacuum states
$\rho_{bj}(0)=|0_j\rangle\langle0_j|$ (in this case, we will
consider $\gamma_j\neq0$) or (ii) thermal equilibrium states
$\rho_{bj}(0)=e^{-H_B/k_BT}/Z$ (in this one, we set $\gamma$ to be
zero to distinguish the effect of temperature from that of
$\gamma$), where $H_B$ is the pure bath part of the whole
Hamiltonian and $Z={\rm Tr}\left(e^{-H_B/k_BT}\right)$ is the
partition function and the Boltzmann constant $k_B$ will be set
to $1$ for the sake of simplicity.\\

For the former case, Eq. \ref{Lind} will be exploited to calculate
$\rho(t)$. For the latter one, Eq. \ref{Lind} is reduced to
\begin{equation}\label{rhot}
\rho(t)=\exp(-iHt)\rho(0)\exp(iHt).
\end{equation}
To determine the dynamics of the density matrix for the whole
system, two factors need to be considered. The first one is the
expression of the thermal bath state. In numerical calculations
\cite{TWmodel}, we have to expand $\rho_{bj}(0)$ ($j=1,2$) to a
summation of its eigenvectors with corresponding weights
determined by its eigenvalues:
\begin{equation}
\rho_{bj}(0)=\sum_m|\phi_{mj}\rangle\omega_{mj}\langle\phi_{mj}|,\quad
\omega_{mj}=\frac{e^{-E_{mj}/T}}{Z_j}
\end{equation}
Then for the two single-modes, we have
\begin{equation}
\rho_{b1}(0)\otimes\rho_{b2}(0)=\sum_{mn}|\phi_{m1}\rangle|\phi_{n2}\rangle
\omega_{mn}\langle\phi_{n2}| \langle\phi_{m1}|,\quad
\omega_{mn}=\frac{e^{-(E_{m1}+E_{n2})/T}}{Z_1Z_2}
\end{equation}
where the subscripts $m$ and $n$ refer to mode $1$ and $2$
respectively. The second important factor is the evaluation of the
evolution operator $U(t)=\exp(iHt)$. A polynomial expansion scheme
proposed by us in Ref. \cite{Jing1, Jing2, Jing3} is applied into
the computation,
\begin{equation}
U(t)=\left(\frac{1}{1+it}\right)^{\alpha+1}
\sum^{\infty}_{k=0}\left(\frac{it}{1+it}\right)^kL^{\alpha}_k(H),
\end{equation}
$L^{\alpha}_k(H)$ is one type of Laguerre polynomials as a
function of $H$, where $\alpha$ ($-1<\alpha<\infty$) distinguishes
different types of the Laguerre polynomials and $k$ is the order
of it. The scheme is of an efficient numerical algorithm motivated
by Ref. \cite{Dobrovitski1, Hu}, which is pretty well suited to
many quantum problems, open or closed. Additionally, it could give
results in a much shorter time compared with the traditional
methods under the same numerical accuracy requirement, such as the
well-known $4$-order Runge-Kutta algorithm. After the density
matrix $\rho(t)$ for the whole system is obtained, the reduced
density matrix $\rho_{S}(t)$ for the two atoms can be derived by
tracing out the degrees of freedom of the two single-mode
cavities.

\section{Simulation results and discussions}\label{discussion}

We discuss three important physical quantities which indicate the
time evolution of the subsystem. (i) The concurrence. It is a very
good measurement for the intra-entanglement between two qubits and
monotone to the quantum entropy of the subsystem when the subsystem
is in a pure state. It is defined as:
\begin{equation}\label{Concurrence}
C=\max\{\lambda_1-\lambda_2-\lambda_3-\lambda_4,~0\},
\end{equation}
where $\lambda_i$ are the square roots of the eigenvalues of the
product matrix
$\rho_S(\sigma^y\otimes\sigma^y)\rho^*_S(\sigma^y\otimes\sigma^y)$
in decreasing order. (ii) The fidelity. It is defined as
\begin{equation}\label{fide}
F(t)={\rm Tr}_S[\rho_{\rm ideal}(t)\rho_S(t)].
\end{equation}
where $\rho_{\rm ideal}(t)$ represents the pure state evolution of
the subsystem only under $H_S$, without interaction with the
environment. In this study,
$H_S=\frac{\omega_1}{2}\sigma^z_1+\frac{\omega_2}{2}\sigma^z_2$. The
fidelity is a measurement for decoherence and depends on
$\rho_{ideal}$. It achieves its maximum value $1$ only if
$\rho_S(t)$ equals to $\rho_{ideal}(t)$. (iii) The entropy exchange
${\rm En}$ is defined as ${\rm En}=-{\rm Tr}(\rho_S\log_2\rho_S)$.
It is the von Neumann entropy of the joint state of the subsystem as
composed of the two qubits in our model. It measures the amount of
the quantum information exchange between the subsystem and the
environment. For the subsystem consisted by two two-level atoms (its
Hilbert space is $4\times4$), the entropy maximum is
$\log_2(4)=2.0$. When it reaches its maximum value, it means all the
quantum information is cast out of the subsystem or the quantum
subsystem degenerates to a classical state.

\subsection{Dynamics at different $\gamma$}\label{discussion:gamma}

\begin{figure}[htbp]
\centering \subfigure[$C(t)$]{\label{ga0011:C}
\includegraphics[width=3in]{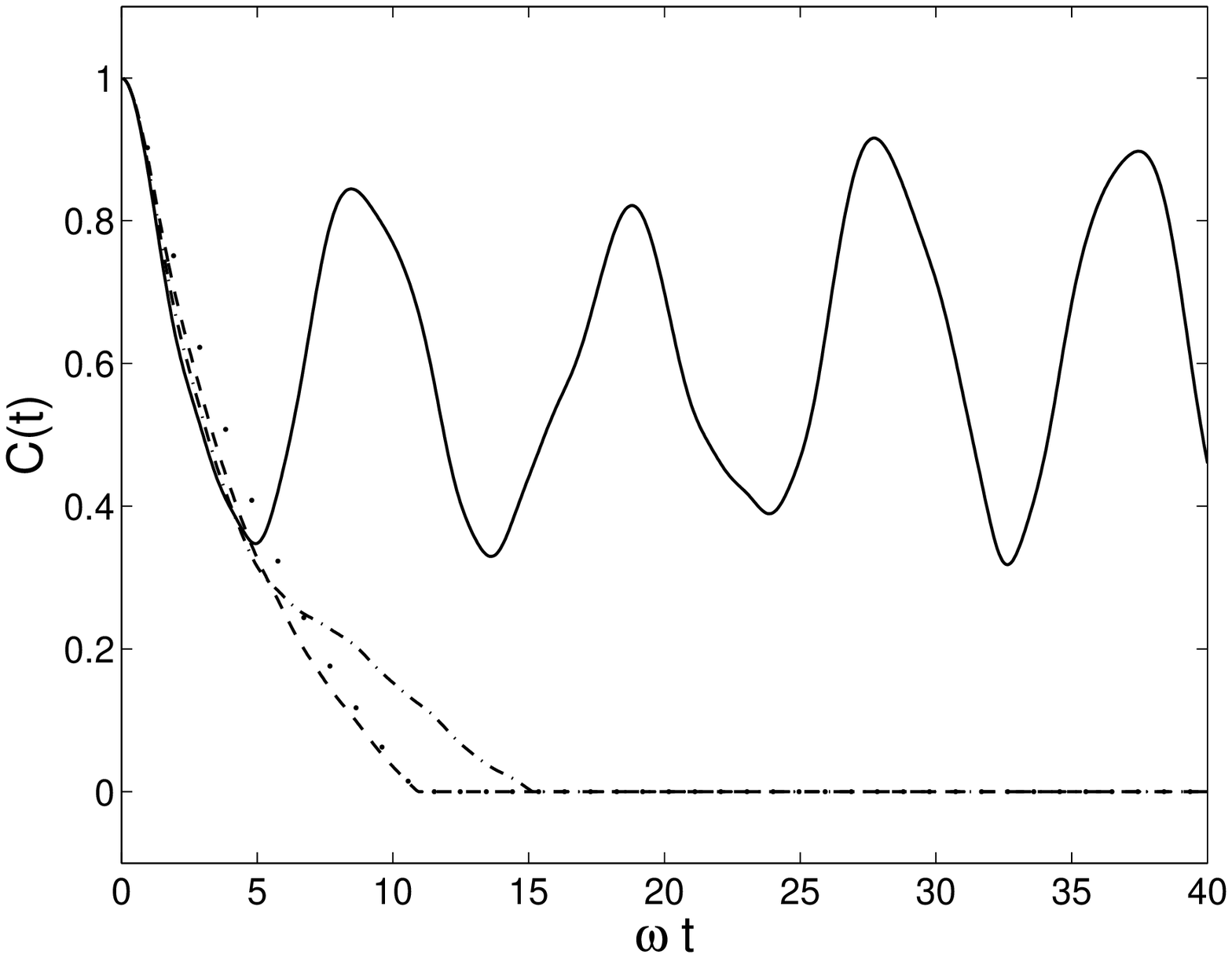}}
\subfigure[$Fd(t)$]{\label{ga0011:Fd}
\includegraphics[width=3in]{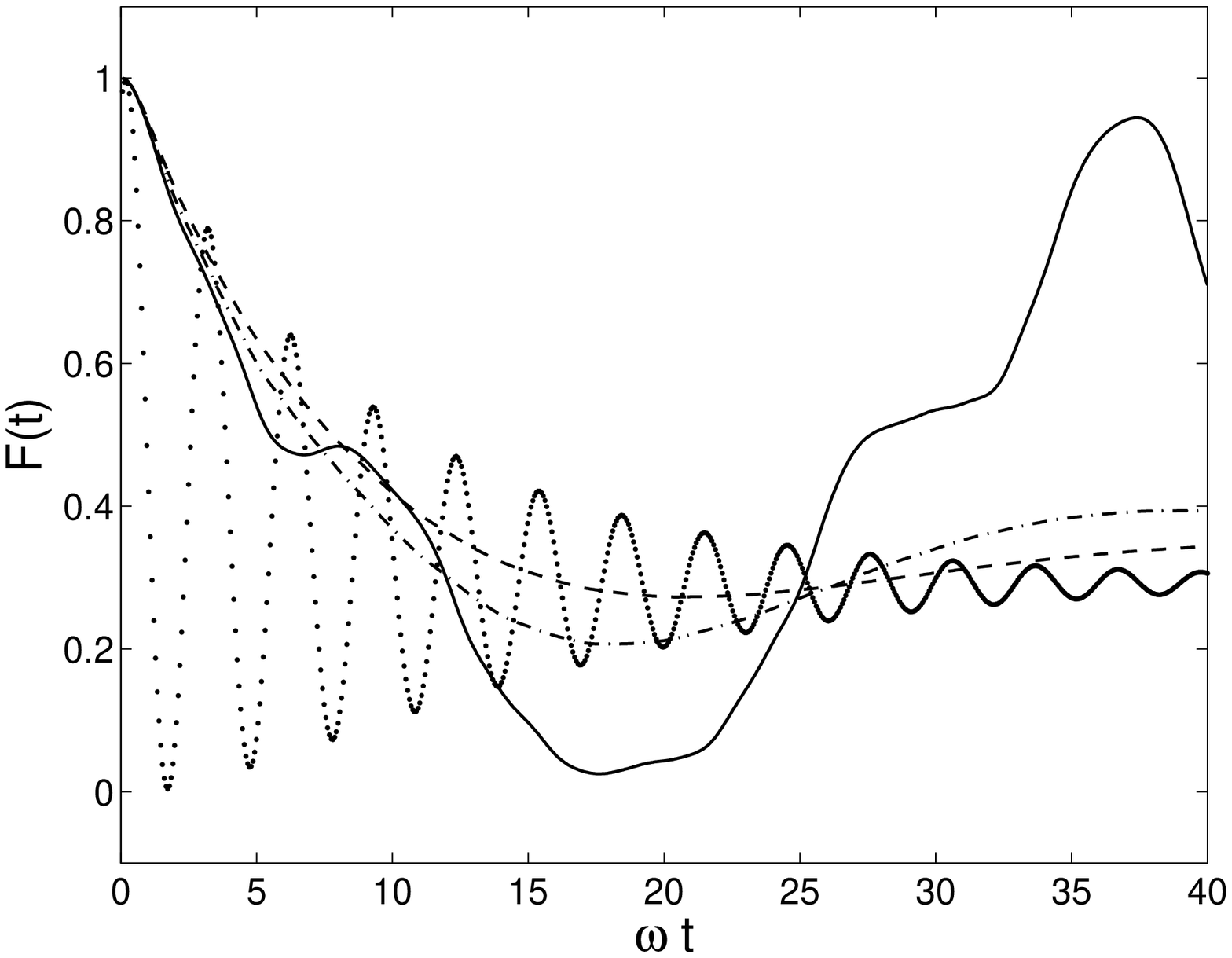}}
\caption{Time evolution for (a) Concurrence, (b) Fidelity with the
subsystem starting from $1/\sqrt{2}(|00\rangle+|11\rangle)$ at
different values of anisotropic parameter: $\gamma=0$ (solid curve),
$\gamma=0.2$ (dot dashed curve), $\gamma=0.4$ (dashed curve),
$\gamma=0.8$ (dotted curve). The two cavities are initialed as
$|0\rangle_1|0\rangle_2$.} \label{ga0011}
\end{figure}

\begin{figure}[htbp]
\centering \subfigure[$C(t)$]{\label{ga0110:C}
\includegraphics[width=3in]{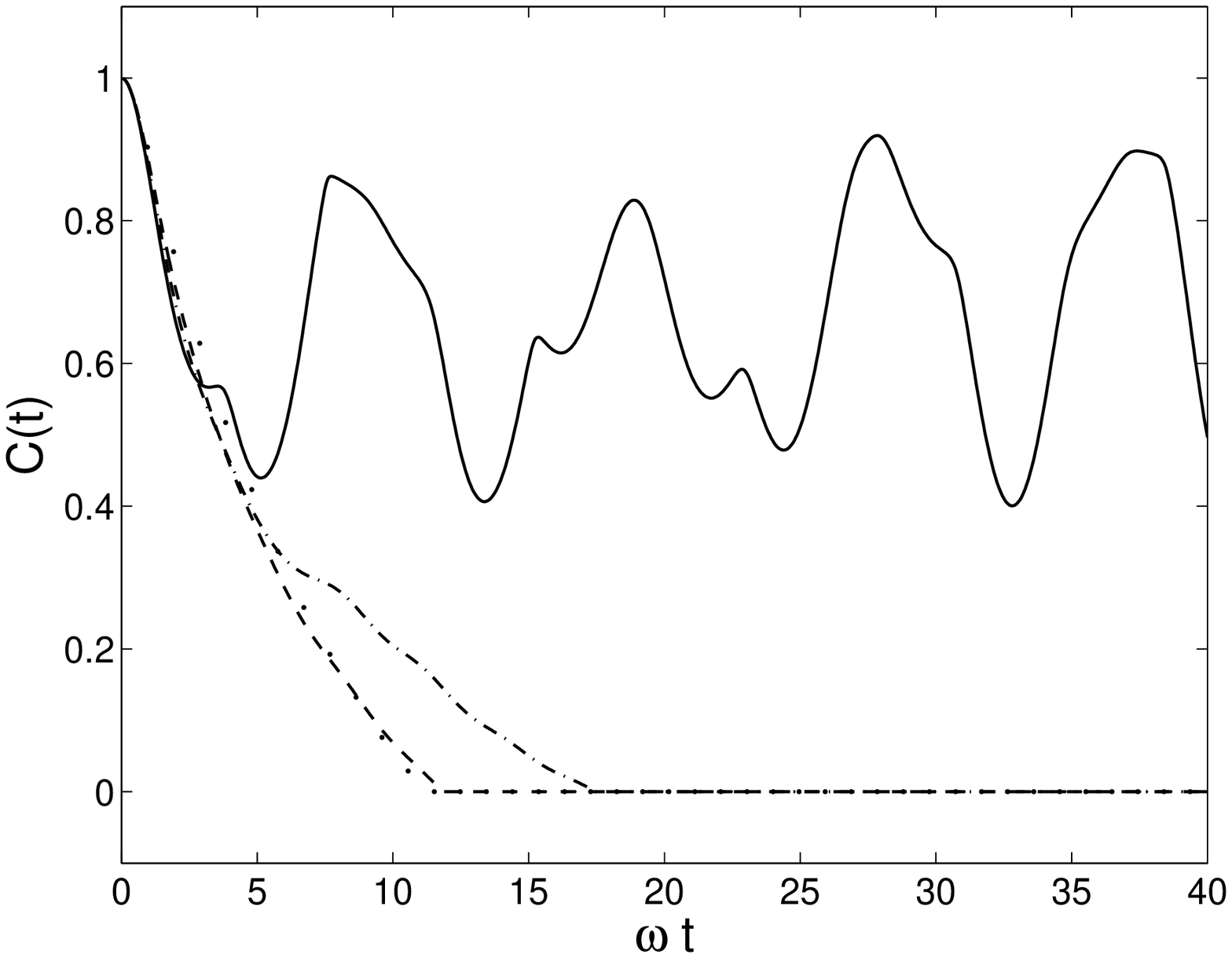}}
\subfigure[$Fd(t)$]{\label{ga0110:Fd}
\includegraphics[width=3in]{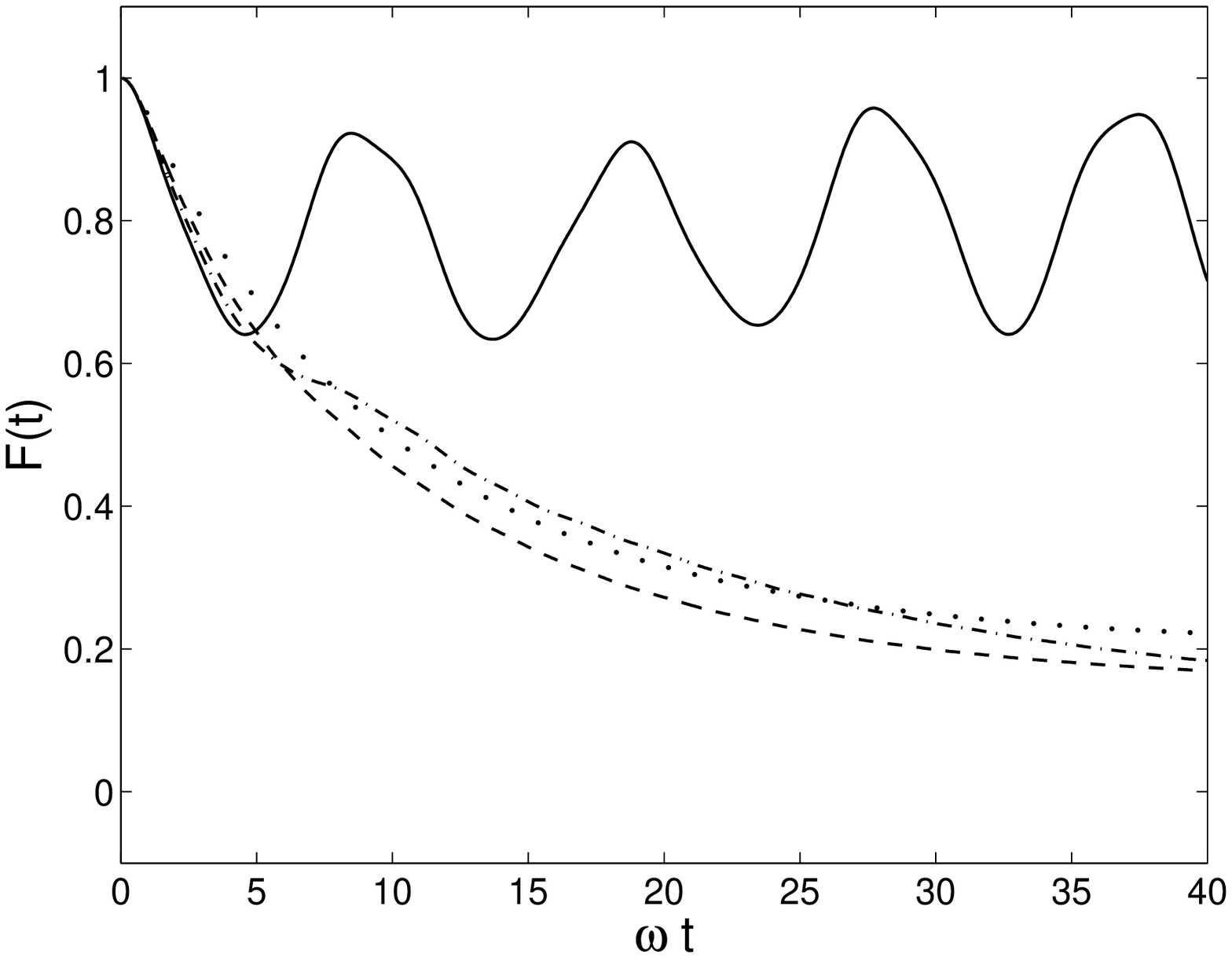}}
\caption{Time evolution for (a) Concurrence, (b) Fidelity with the
subsystem starting from $1/\sqrt{2}(|01\rangle+|10\rangle)$ at
different values of anisotropic parameter: $\gamma=0$ (solid curve),
$\gamma=0.2$ (dot dashed curve), $\gamma=0.4$ (dashed curve),
$\gamma=0.8$ (dotted curve). The two cavities are initialed as
$|0\rangle_1|0\rangle_2$.} \label{ga0110}
\end{figure}

\begin{figure}[htbp]
\centering
\subfigure[$1/\sqrt{2}(|01\rangle+|10\rangle)$]{\label{gaEn:0110}
\includegraphics[width=3in]{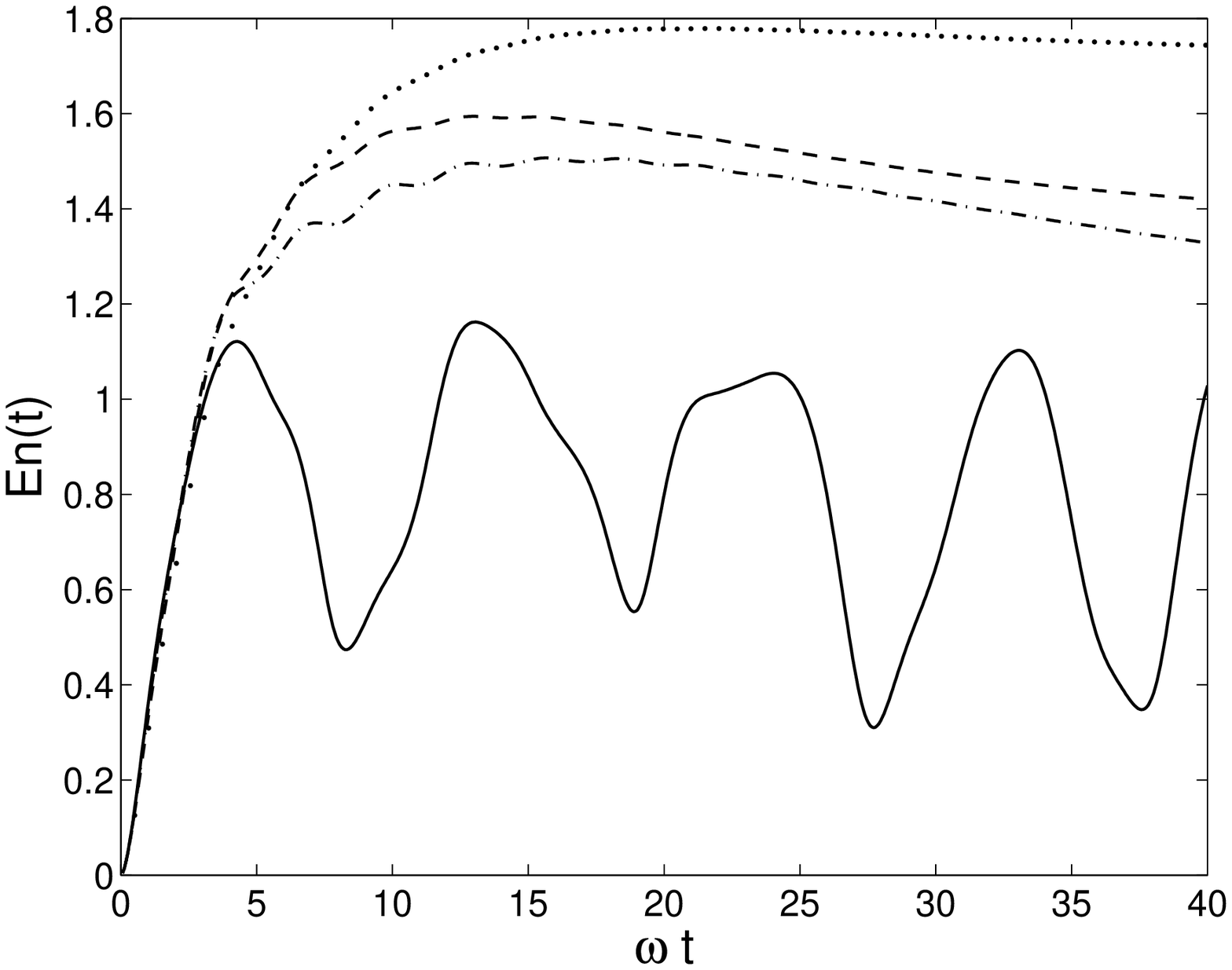}}
\subfigure[$1/\sqrt{2}(|00\rangle+|11\rangle)$]{\label{gaEn:0011}
\includegraphics[width=3in]{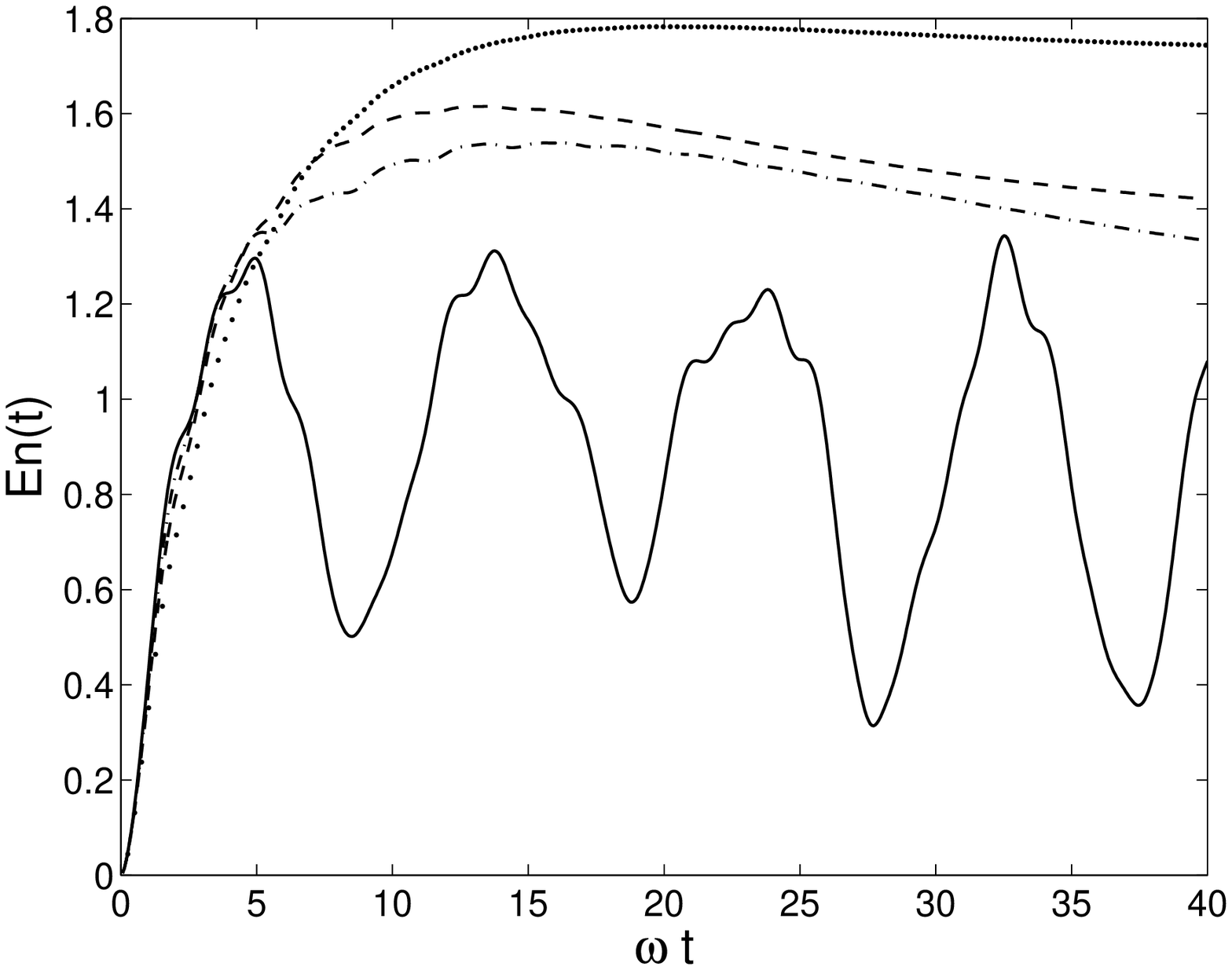}}
\caption{Time evolution for entropy of the subsystem from (a)
$1/\sqrt{2}(|01\rangle+|10\rangle)$, (b)
$1/\sqrt{2}(|00\rangle+|11\rangle)$ at different values of
anisotropic parameter: $\gamma=0$ (solid curve), $\gamma=0.2$ (dot
dashed curve), $\gamma=0.4$ (dashed curve), $\gamma=0.8$ (dotted
curve). The two cavities are initialed as $|0\rangle_1|0\rangle_2$.}
\label{gaEn}
\end{figure}

In order to discuss the effect of $\gamma$ (We suppose the two
cavities have the same loss degree: $\gamma_1=\gamma_2=\gamma$.) and
$T$, all the other parameters are fixed for the sake of simplicity
and without loss of generality:
\begin{eqnarray*}
\omega_1&=&\omega_2=\omega=0.4, \\
\epsilon_1&=&\epsilon_2=\epsilon=-0.5,\\
g_1&=&g_2=g=0.2.
\end{eqnarray*}
And we choose $1/\sqrt{2}(|00\rangle+|11\rangle)$ and
$1/\sqrt{2}(|01\rangle+|10\rangle)$ as two different initial states
for the subsystem.\\

We first discuss the effect of the photon loss rate $\gamma$. It is
evident that with a larger $\gamma$, the entanglement degree of the
subsystem will decrease in a faster speed, which could be verified
by Fig. \ref{ga0011:C} and Fig. \ref{ga0110:C}. The tendency of the
two cases is similar. If $\gamma=0$, the concurrence will oscillate
periodically with time and will not be dissipated; but the peak
value of it will never reach $1.0$. On the whole, the curves of the
concurrence are not perfectly harmonic, which is a little different
with the results gotten in previous works. It is due to the
stochastic and irrelevant microscopical processes (the spins drop
from the excited state by emitting a photon or jump to the excited
state by absorbing a phonon) inside the two different cavities. And
the dynamics of the concurrence stems from such numerous processes,
so the evolution is approximately harmonic but not perfect. When
$\gamma>0$, the concurrence drops abruptly to zero in a short time.
It coincides with the description about entanglement sudden death
(ESD) in Ref. \cite{Yu3, Yu4} that ``after the concurrence goes
abruptly to zero, it arises more or less from nowhere''. This is an
example of ESD. The photons leaking out of the cavities greatly
reduced the nonlocal connection between the two qubits. When
$\gamma$ is bigger than $0.4$, the speed of ESD is saturated and we
almost cannot distinguish the curve of $\gamma=0.4$ from that of
$\gamma=0.8$. The state $1/\sqrt{2}(|01\rangle+|10\rangle)$ seems
more robust than $1/\sqrt{2}(|00\rangle+|11\rangle)$. It is verified
that, for example, in the condition of $\gamma=0.2$, the concurrence
of the former state (to see the dot dashed line in Fig.
\ref{ga0110:C}) decreases to zero at $\omega t=17.472$ for the first
time, while the latter one does at $\omega t=15.168$. \\

The fidelity dynamics of the two Bell states, however, is much
different from each other. In Fig. \ref{ga0011:Fd}, when there is no
cavity leakage, the fidelity is not synchronous with the
concurrence. Concretely, along that curve, only when $\omega
t=37.376$, most component of the subsystem state is recovered to its
initial one with high fidelity $F(t)=0.944346$. While the other
three peaks inside the interval of $\omega t\in[5.0, 35.0]$ along
the solid line in Fig. \ref{ga0011:C} are just fake phenomena:
although the entanglement degree between the two subsystem atoms is
high, but the state of the subsystem is different from the initial
one. The tendency of $\gamma=0.2$ and $\gamma=0.4$ is in a similar
manner: the curves decrease smoothly and move towards
$F=0.3\sim0.4$. While the curve of $\gamma=0.8$ has an oscillating
dynamics with gradual shrinking amplitude. Finally, its value
approaches to $F=0.35$. In Fig. \ref{ga0110:Fd}, however, when
$\gamma=0$, the fidelity is synchronous with the concurrence. Thus
the fidelity of the state $|01\rangle+|10\rangle$ is more robust
than that of $|00\rangle+|11\rangle$ under the same environment.
These contrasts between the two initial state can be noticed in
other works. It implies the physics essence of them is different,
although both of them are of the most entangled states. Yet the
other three cases with leakage $\gamma>0$ have almost the same kind
of dynamics for the two states. The descend speed of the fidelity
decreases with time and the value of fidelity approaches to about
$F=0.2$ at $\omega t=40.0$. \\

The quantum evolution of entropy exchange $\rm{En}$ is depicted in
Fig. \ref{gaEn}. Initially, the entropy of the two entangled qubits
equals to $0$, which means that all the quantum information is kept
in the entanglement between the two qubits. We find that when
$\gamma=0$, the evolutions of the two Bell states are a little
different but the oscillation periods of them are almost the same.
Although there is no leakage, but the bath, the two single-mode
cavities, will absorb some of the quantum information inside the
subsystem, which means the entropy never goes back to zero as
initialled. To compare the solid line in Fig. \ref{gaEn:0011} (Fig.
\ref{gaEn:0110}) with that in Fig. \ref{ga0011:C} (Fig.
\ref{ga0110:C}), we notice that the tendency of the entropy is
opposite to that of the concurrence. The dynamics difference from
the two initial states can almost be removed by introducing non-zero
$\gamma$ as it is shown in the comparison of Fig. \ref{gaEn:0110}
and Fig. \ref{gaEn:0011}. It is evident that with a larger $\gamma$,
more quantum information of the subsystem is transferred into the
bath.

\subsection{Dynamics under different $T$}

\begin{figure}[htbp]
\centering \subfigure[$C(t)$]{\label{T0011:C}
\includegraphics[width=3in]{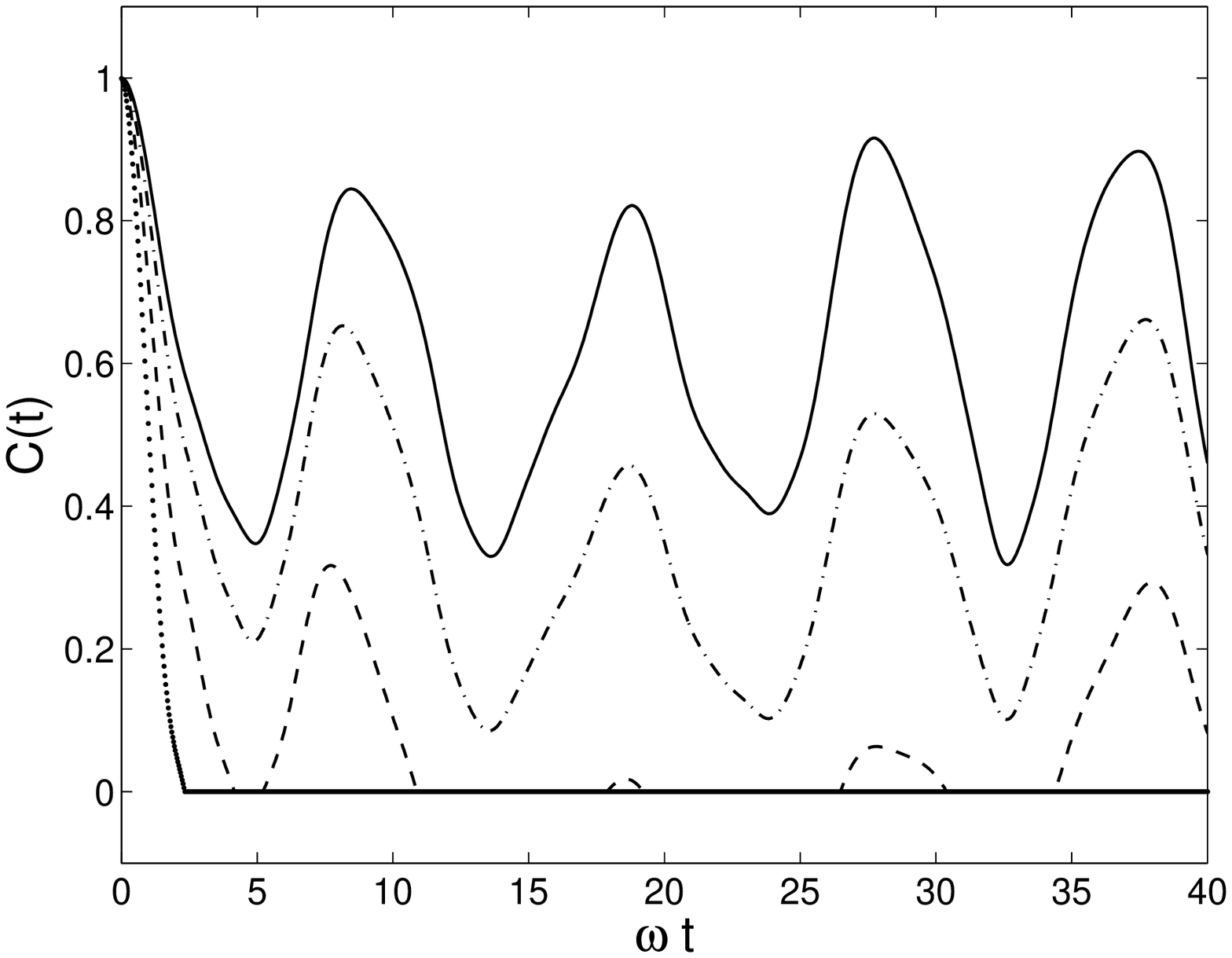}}
\subfigure[$Fd(t)$]{\label{T0011:Fd}
\includegraphics[width=3in]{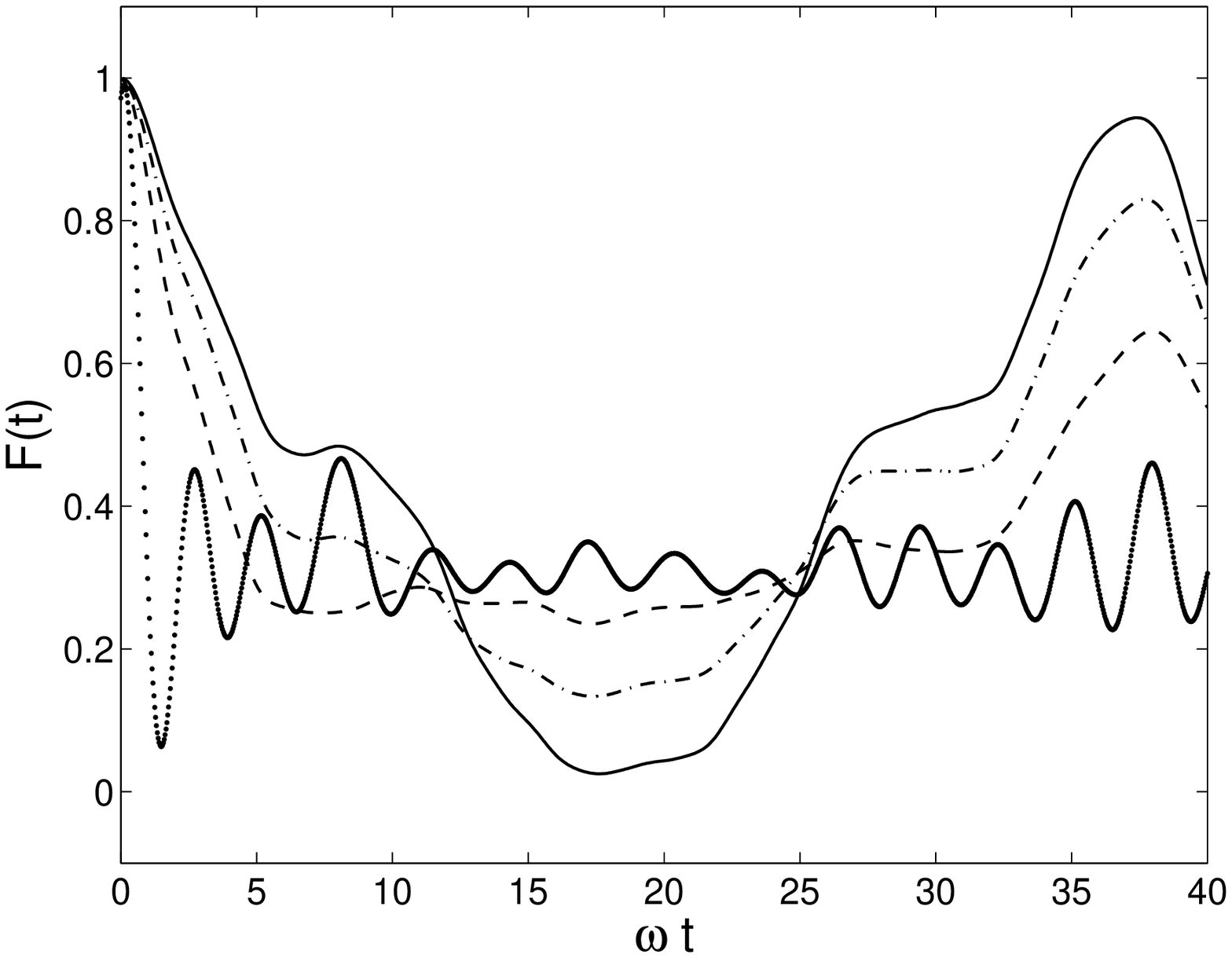}}
\caption{Time evolution for (a) Concurrence, (b) Fidelity with the
subsystem starting from $1/\sqrt{2}(|00\rangle+|11\rangle)$. The two
cavities are initialed in thermal states at different temperature:
$T=0$ (solid curve), $T=0.25\omega$ (dot dashed curve),
$T=0.5\omega$ (dashed curve), $T=1.0\omega$ (dotted curve).}
\label{T0011}
\end{figure}

\begin{figure}[htbp]
\centering \subfigure[$C(t)$]{\label{T0110:C}
\includegraphics[width=3in]{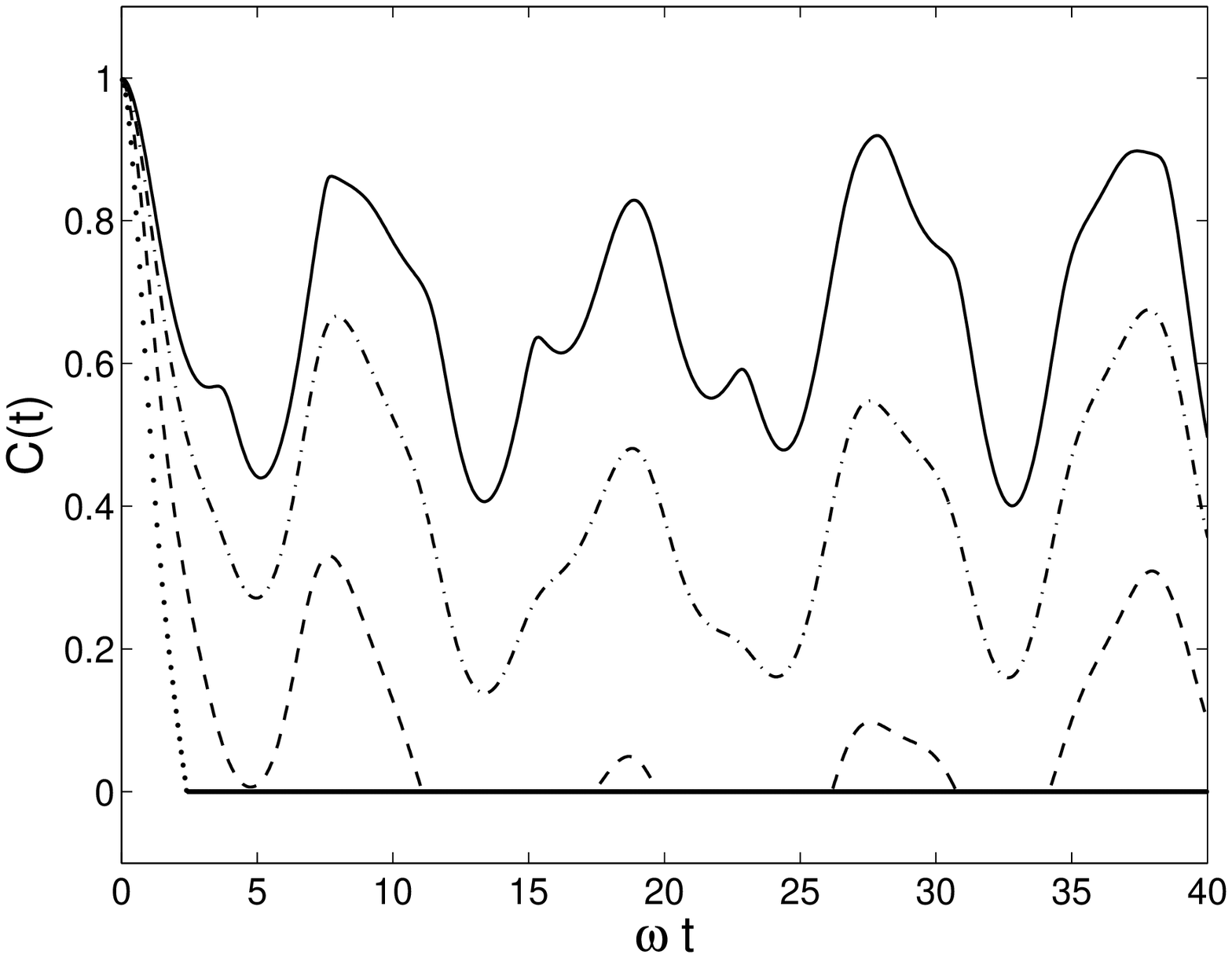}}
\subfigure[$Fd(t)$]{\label{T0110:Fd}
\includegraphics[width=3in]{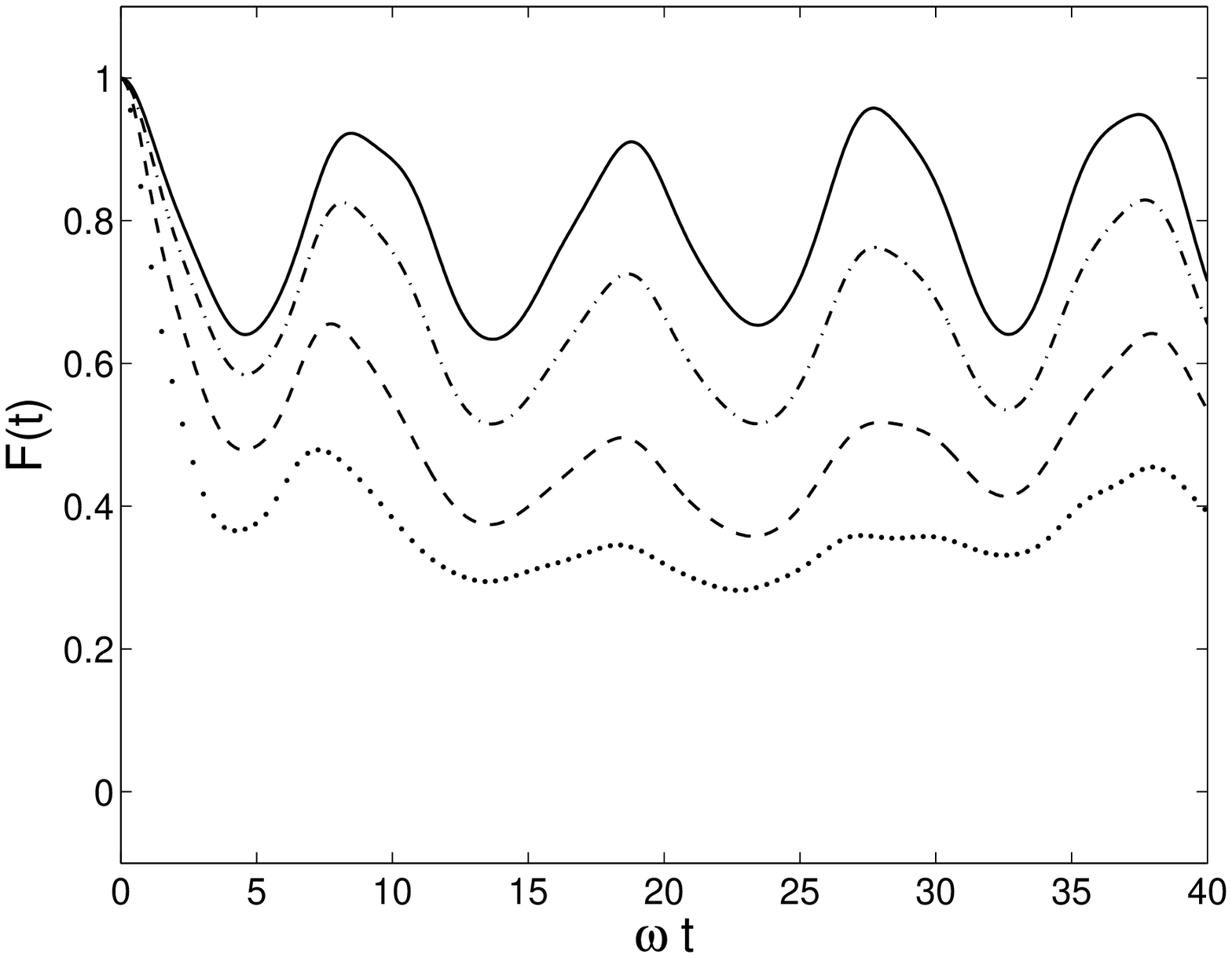}}
\caption{Time evolution for (a) Concurrence, (b) Fidelity with the
subsystem starting from $1/\sqrt{2}(|01\rangle+|10\rangle)$. The two
cavities are initialed in thermal states at different temperature:
$T=0$ (solid curve), $T=0.25\omega$ (dot dashed curve),
$T=0.5\omega$ (dashed curve), $T=1.0\omega$ (dotted curve).}
\label{T0110}
\end{figure}

\begin{figure}[htbp]
\centering
\subfigure[$1/\sqrt{2}(|01\rangle+|10\rangle)$]{\label{TEn:0110}
\includegraphics[width=3in]{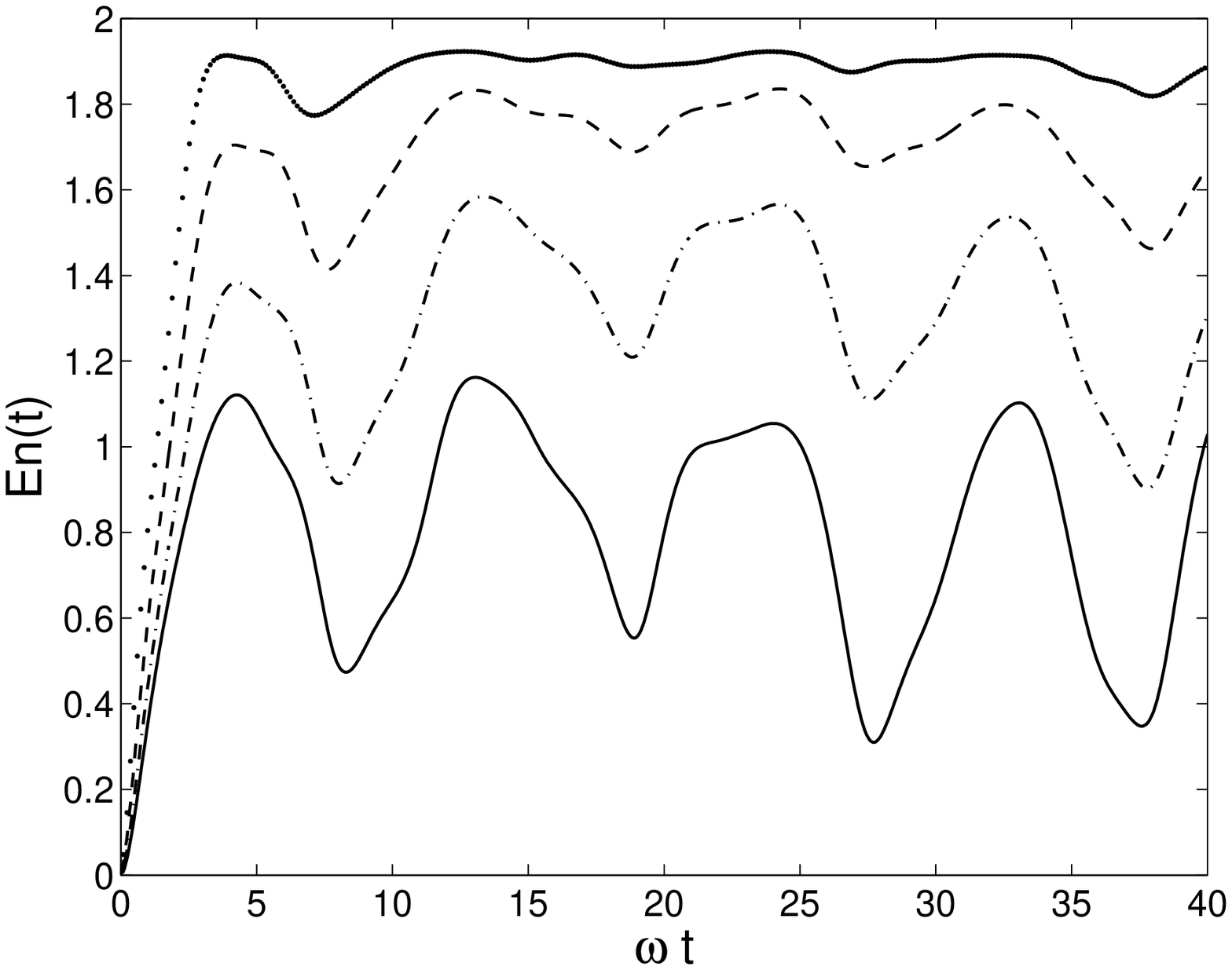}}
\subfigure[$1/\sqrt{2}(|00\rangle+|11\rangle)$]{\label{TEn:0011}
\includegraphics[width=3in]{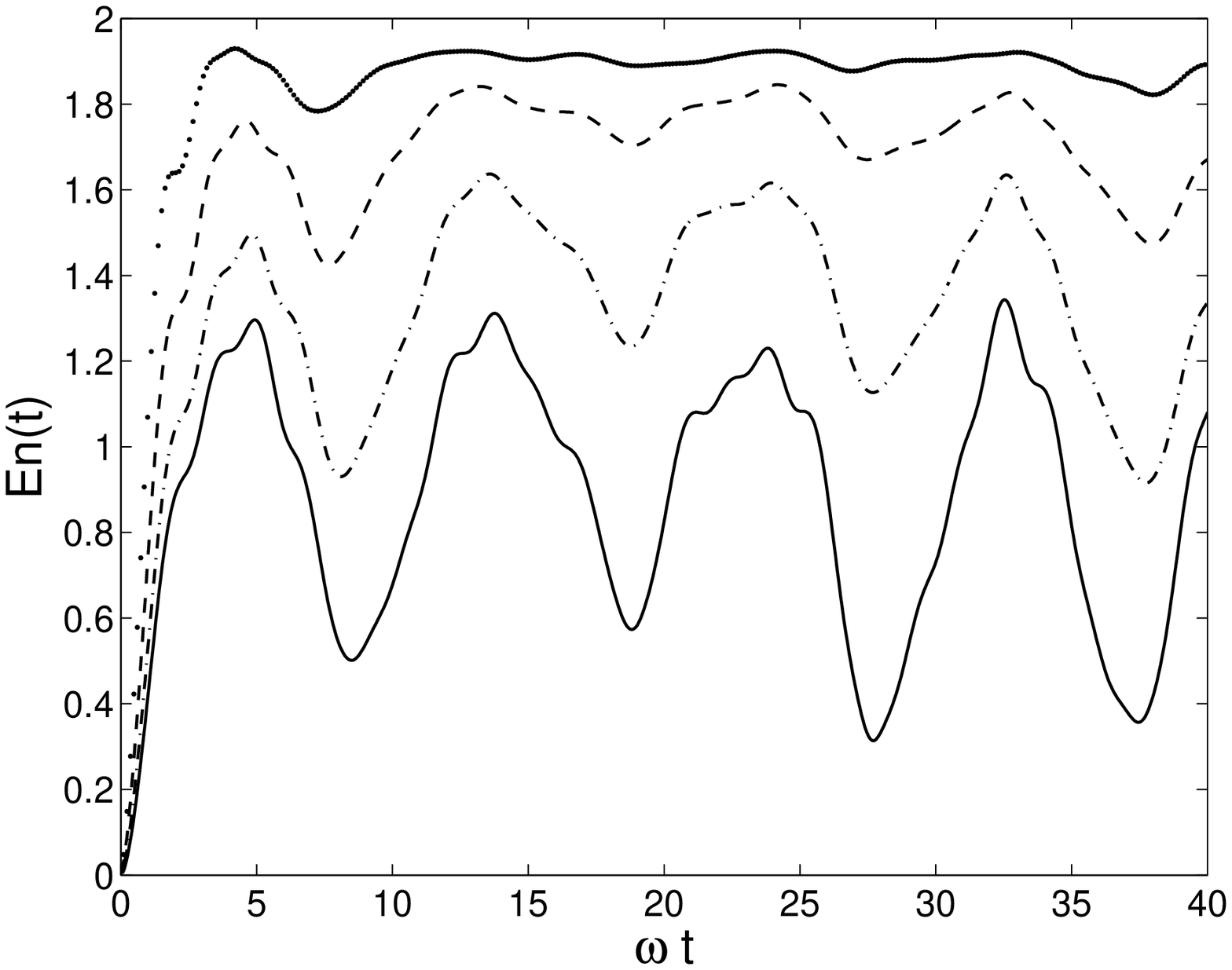}}
\caption{Time evolution for entropy of the subsystem from (a)
$1/\sqrt{2}(|01\rangle+|10\rangle)$, (b)
$1/\sqrt{2}(|00\rangle+|11\rangle)$ on the condition that the two
cavities are initialed in thermal states at different values of
temperature: $T=0$ (solid curve), $T=0.25\omega$ (dot dashed
curve), $T=0.5\omega$ (dashed curve), $T=1.0\omega$ (dotted
curve).} \label{TEn}
\end{figure}

In this subsection, we turn to the effect of temperature of the
cavity modes. It is found that when the temperature is comparatively
low, $T=0.25\omega$, the entanglement degree oscillates but will not
corrupt to a sudden death (to see the dot dashed line in Fig.
\ref{T0011:C} and Fig. \ref{T0110:C}). When it goes up to a moderate
temperature $T=0.5\omega$, the ESD happens. The first moment at
which the concurrence decreases to zero and does not revive
immediately is dependent on the initial state. For
$1/\sqrt{2}(|00\rangle+|11\rangle)$, it takes place at $\omega
t=4.192$, which is much earlier than that of
$1/\sqrt{2}(|01\rangle+|10\rangle)$, $\omega t=11.104$. This
difference coincides with the comparison in subsection
\ref{discussion:gamma}, which means the entanglement damping speed
of the state $|01\rangle+|10\rangle$ is slower than that of state
$|00\rangle+|11\rangle)$. It is hinted that the bath influence on
the former state is comparatively inapparent. Yet both of them can
still revive to a certain extent $0.2 \sim 0.3$ after some time.
When the temperature increases to $T\geq0.75 \omega$, it can not
revive in the future after the first sudden death happens for both
Bell states. Till the temperature is as high as $T=1.0 \omega$, the
concurrence of both cases falls with a very quick speed to zero.
Obviously, the temperature will destroy the initial most-entangled
states even if it is much lower than the energy bias $\omega$ of the
two-level atoms. After ESD happened, the quantum oscillation from
the local thermal bath may help to entangle the two qubits, but this
positive effect is neglectable when the temperature is high enough.
Then the entanglement between the Bell states is damped forever. \\

Under variant temperature, the different dynamics of fidelity, which
rely on the initial state, are shown in Fig. \ref{T0011:Fd} and Fig.
\ref{T0110:Fd}. In a short interval after $\omega t=0$, higher
temperature means faster damp speed for both initial states. Yet in
a long time scale, their actions are totally different. For
$T\leq0.5\omega$, the four curves evolves pseudo-periodically, but
the period of $1/\sqrt{2}(|00\rangle+|11\rangle)$ is much larger
than that of $1/\sqrt{2}(|01\rangle+|01\rangle)$. In Fig.
\ref{T0011:Fd}, the dotted curve of $T=1.0\omega$ fluctuates with
time, whose amplitude damps from the beginning time and gets some
revival when $\omega t>25.0$. In Fig. \ref{T0110:Fd}, the
oscillation evolution at temperature $T=1.0\omega$ is in the same
manner as those at $T=0.25\omega$ and $T=0.5\omega$. And their
amplitudes and peak values of the fidelity decrease with increasing
temperature. Obviously, an environment at higher temperature
destroys the fidelity of the subsystem even stronger. \\

In Fig. \ref{TEn}, we give the dynamics of the entropy exchange
under different temperatures. Although there are some differences
between the two sub-figures even when $T$ is as high as $0.5\omega$,
but the tendency of the two cases are almost the same. With higher
$T$, the entropy increases with faster speed and behaves an
oscillation evolution with smaller amplitude. It is important to
find that the entropy exchange in Fig. \ref{TEn:0110} exhibits an
opposite behavior in comparison with that of the fidelity in Fig.
\ref{T0110:Fd}. The periods of the two evolutions are the same and
when the fidelity experiences a peak value, (For instance, for the
dot dashed curves in the two figures, when $T=0.25\omega$, the four
peaks appear at $\omega t=8.160$, $\omega t=18.656$, $\omega
t=27.776$ and $\omega t=37.696$ in the given interval) the entropy
is at the corresponding valley point and vice versa. In Ref.
\cite{Xiong}, the authors found the entropy exchange exhibit the
behavior opposite to that of the concurrence. That coincides with
our results. In our model, when the temperature is not too high and
the initial Bell state is $1/\sqrt{2}(|01\rangle+|10\rangle)$, the
concurrence and fidelity evolve synchronously, then the dynamics of
entropy is also opposite to that of the concurrence. However, when
the temperature is high enough, the ESD will make this
synchronization relationship invisible. So the opposite relationship
between the entropy and concurrence is lost. Therefore we have to
conclude that the more quantum information of the subsystem
transfers to its bath, the more fidelity of that decreases during
the time interval if
$|\psi(0)\rangle=1/\sqrt{2}(|01\rangle+|10\rangle)$.

\section{Conclusion}\label{conclusion}

In conclusion, we investigate the dynamics of two distinct uncoupled
qubits embedded respectively in two single-mode cavities with
leakage, which constitute the environment in our model. The
subsystem consisting of the two qubits is initially prepared as one
of the Bell states and the cavities as vacuum states or thermal
equilibrium states. Under these two situations, the concurrence, the
fidelity and the entropy exchange are used to portrait the subsystem
dynamics from different views. Polynomial expansion method is
applied into the numerical calculation. It is found that (i) for the
leaky cavities, when the loss rate $\gamma\geq0.4$, the speed of
entanglement sudden death achieves its maximal value; (ii) The
evolution of different Bell states can easily be distinguished by
their fidelity dynamics; (iii) the behavior of entropy exchange is
opposite to that of the fidelity and concurrence in certain
conditions.

\begin{acknowledgments}
We would like to acknowledge the support from the National Natural
Science Foundation of China under grant No. 10575068, the Natural
Science Foundation of Shanghai Municipal Science Technology
Commission under grant Nos. 04ZR14059 and 04dz05905 and the CAS
Knowledge Innovation Project Nos. KJcx.syw.N2.

\end{acknowledgments}

\end{document}